\documentclass[12pt,preprint]{aastex}
\shorttitle{}
\shortauthors{Nesvorn\'y}
\begin{document}
\title{Dynamical Origin and Terrestrial Impact Flux \\of Large Near-Earth Asteroids}
\author{David Nesvorn\'y$^{1,2}$, Fernando Roig$^2$}
\affil{Department of Space Studies, Southwest Research Institute, 1050 Walnut St., \\Suite 300, 
Boulder, CO 80302, USA} 
\affil{Observat\'orio Nacional, Rua Gal. Jose Cristino 77, Rio de Janeiro, RJ 20921-400, Brazil}
\begin{abstract}
Dynamical models of the asteroid delivery from the main belt suggest that the current impact 
flux of diameter $D>10$ km asteroids on the Earth is $\simeq0.5$-1~Gyr$^{-1}$. Studies of the 
Near-Earth Asteroid (NEA) population find a much higher flux, with $\simeq$7 $D>10$-km asteroid 
impacts per Gyr. Here we show that this problem is rooted in the application of impact 
probability of small NEAs ($\simeq1.5$ Gyr$^{-1}$ per object), whose population is well 
characterized, to large NEAs. In reality, large NEAs evolve from the main belt by different 
escape routes, have a different orbital distribution, and lower impact probabilities ($0.8\pm0.3$ 
Gyr$^{-1}$ per object) than small NEAs. In addition, we find that the current population 
of two $D>10$ km NEAs (Ganymed and Eros) is a slight fluctuation over the long term average of 
$1.1\pm0.5$ $D>10$ km NEAs in a steady state. These results have important implications for 
our understanding of the occurrence of the K/T-scale impacts on the terrestrial worlds.     
\end{abstract}
\keywords{minor planets, asteroids: general}
\section{Introduction}
The impact cratering record of the Moon and terrestrial planets provides important clues about 
the impactor flux in the inner Solar System. The number of craters recorded on each world's surface is a 
measure of the intensity of the bombardment that the surface experienced over its age. If the surface 
age is known, the crater data can be interpreted to yield clues about the nature of the impactor 
population. If the impactor population is well characterized from independent means, the crater data can 
be used to estimate the surface age. It is often the case, however, that both the surface age and the 
properties of the impactor population are uncertain, leaving us with a difficult task mathematically 
equivalent to one equation with two unknowns. Things are further complicated by uncertainties in the 
scaling laws that are used to calculate the crater size from the impactor's mass and velocity
(e.g., Holsapple 1993, Johnson et al. 2016). 

Historically, the problem of interpreting crater records on different worlds has been 
approached from two different angles. In the first approach, which is firmly set in the realm of 
{\bf data-driven} science, the crater counts are obtained from the existing imagery and used to 
define the so-called {\it production function}, which, apart from uncertainties in the scaling 
laws, conveys the size and time dependence of the impactor population (e.g., Neukum et al. 2001,
Hartmann \& Neukum 2001, Marchi et al. 2012, Robbins 2014). The results of crater counts are often correlated 
among different worlds to demonstrate the applicability, or the lack of it, of the production function 
in different parts of the Solar System, and during different epochs of the Solar System history. 

The second, {\bf theory-driven} approach strives to understand the impactor population by modeling the 
orbital evolution and estimating the impact probabilities (e.g., Bottke et al. 2002, 2012; Morbidelli 
et al. 2002; Minton \& Malhotra 2010; Nesvorn\'y et al. 2017). A theoretical model expresses our expectations 
about the impact flux on different worlds and its time variability. It is often calibrated on telescopic 
observations of small-body populations (e.g., near-Earth asteroids), and thus represents a useful link between
different datasets (e.g., between crater counts and telescopic observations of NEAs and main belt asteroids). 

Each of the methods described above has its shortcomings. The data-driven approach yields a crater 
production function but does not tell us, at least not immediately, the {\it meaning} of the cratering
record in the context of the known Solar System architecture (e.g., the source of impactors). 
The modeling approach, on the other hand, can struggle to accurately represent the reality, because the 
populations of small bodies may not be well characterized from observations, especially at small sizes, or 
because the computer simulations may not be accurate (poor resolution due to CPU limitations, uncertain 
nature of dynamical processes, etc.). These and other issues often lead to a situation where inconsistencies 
arise between the model expectations and data-based inferences about the impactor flux. 
\section{Motivating Problem}
Here we aim at resolving one such inconsistency related to the present-day impact flux 
in the inner Solar System. Stated briefly, the problem arises when one compares the impact flux inferred 
from analysis of NEAs with the impact flux obtained from dynamical models 
of NEA delivery from the main asteroid belt. For example, Johnson et al. (2016) reported, adopting the 
former approach and assuming that the NEA impact flux was constant over the past 1~Gyr, that 
$\simeq7$ impacts of $D > 10$-km asteroids should have occurred on the Earth in the last 1 Gyr.
A similar value was previously reported in many other publications, including Chapman \& 
Morrison (1994), Stuart \& Binzel (2004), Le Feuvre \& Wieczorek (2011) and Harris \& D'Abramov (2015).
In contrast, Nesvorn\'y et al. (2017) estimated from their dynamical modeling that only $\sim$0.5-1 such 
impacts should occur. The model impact rates are therefore $\sim$7-14 times below expectations. A similar 
discrepancy was noted in Minton \& Malhotra (2010), who compared their model results with Neukum's 
production function (Neukum et al. 2001). 

This can mean one of several things. First, the current-day impactor flux may be, for some reason, 
larger than the average flux over the past Gyr (e.g., Culler et al. 2000, Mazrouei et al. 2017). 
Alternatively, some model parameters in Nesvorn\'y et al. (2017) and Minton \& Malhotra (2010) may need to 
be tweaked to increase the 
computed impact flux in the last 1 Gyr. It is not clear, however, how this can be done, because the impact 
flux in the last 1 Gyr was shown to be nearly independent, to within a factor of $\sim$2, of various model 
parameters (such as history of planetary orbits, distribution of main belt asteroids, etc.; see discussion 
in Nesvorn\'y et al. 2017). The uncertainties in the model thus do not appear to be responsible for 
the identified problem. Here we demonstrate, instead, that the problem appears because the NEA-based 
estimate of the impact flux of {\it large} asteroids in the works cited above is inaccurate.

To estimate the impact rate, the NEA-based studies assumed that the impact probability of large NEAs is 
essentially the same as the impact probability of small NEAs. Consequently, they used an average 
value for the NEA impact probability, and the size distribution of NEAs, to compute the current impact flux 
of asteroids with $1<D<10$ km. The impact flux was extrapolated to $D>10$ km using the size distribution
of main belt asteroids. This approach is adequate for impactors with $D \sim 1$ km, where the NEA 
size and orbital distributions are well characterized. It may not be fully appropriate, however, 
to extrapolate the results to large asteroids, because it is not guaranteed that the orbital distribution
of large NEAs and their impact probabilities are the same (see, e.g., Valsecchi \& Gronchi 2011).

On one hand, large asteroids typically reach the NEA orbits via slow diffusion in weak resonances 
(Migliorini et al. 1998, Morbidelli \& Nesvorn\'y 1999, Farinella \& Vokrouhlick\'y 1999). The small asteroids, 
on the other hand, can drift over a considerable radial distance by the Yarkovsky effect and reach NEA space 
from the powerful $\nu_6$ resonance at the inner edge of the asteroid belt (Bottke et al. 2002). The $\nu_6$ 
resonance is known to produce highly evolved NEA orbits and impact probabilities on Earth in excess of 1\% 
(Gladman et al. 1997). The small NEAs can therefore have larger impact probabilities than the large ones 
(Valsecchi \& Gronchi 2015). If correct, this would imply that the current impact flux should drop more 
steeply with impactor size (see, e.g., Fig. 1 in Johnson et al. (2016) or Fig. 5 in Harris \& D'Abramov 
(2015)), therefore implying fewer than $\simeq7$ $D>10$-km asteroid impacts in the last 1 Gyr. 

It is difficult to test this conjecture based on the NEA population alone, because there are not enough  
$D>10$ km NEAs, and the statistical inferences based on them would not produce robust results. The 
dynamical model of Nesvorn\'y et al. (2017) also cannot be used to resolve this problem with confidence, because several
model approximations make it a less than ideal gauge of the current impact flux (the model was developed 
to study the dependence of the {\it historical} impact flux of planetary impactors under different assumptions).     
For example, the main belt structure was not exactly reproduced in Nesvorn\'y et al. (2017). In addition, 
their simulations neglected the Yarkovsky effect. According to Bottke et al. (2006), the maximum Yarkovsky 
drift rate of a $D=10$-km inner-belt asteroid is $\simeq2\times10^{-5}$ au Myr$^{-1}$. Thus, the expected 
maximum drift over 1 Gyr is roughly 0.02 au. This seems small but should be significant, because weak 
diffusive resonances in the inner main belt, which provide the main escape routes for large asteroids, 
are very dense, and the nearest one may be only a tiny fraction of an au away. It is therefore important 
to include the Yarkovsky effect to do things correctly.

The Yarkovsky effect also appears to be responsible for the related problem reported in Minton \& Malhotra 
(2010). They performed simulations of the main belt asteroids without the Yarkovsky effect, calibrated their
model on the number of large asteroids remaining in the belt today, and found that on average 
$\sim$1 $D>10$ km asteroids should have impacted on Earth in the last Gyr. Note that this estimate 
is consistent, to within a factor of 2, with the one derived from a similar model in Nesvorn\'y et al. (2017), 
but $\sim$7 times lower than the one reported in Johnson et al. (2016). The reason why they obtained 
a slightly higher impact flux than in Nesvorn\'y et al. (2017) is probably related to the fact that
Minton \& Malhotra (2010) did not account for the early planetary migration/instability which acted 
to deplete the inner part ($a<2.5$ au) of the main belt (Nesvorn\'y et al. 2017). Their model may thus 
slightly overestimate the impact flux, because the inner belt, which produces most NEA impactors, is more 
populated in their model than in reality (see Figure 4 in Minton \& Malhotra 2010). 

Minton \& Malhotra (2010) compared the impact flux obtained in their work with the Neukum production function 
(Neukum et al. 2001) and found that, when the production function is applied to large impactors, it 
predicts $\simeq$6-18 terrestrial impacts of $D>10$~km asteroids in the last 1~Gyr. Neukum and collaborators, 
however, developed their production function mainly from the analysis of counts of small recent craters that 
were produced by $D \ll 10$~km projectiles, for which the Yarkovsky effect is very important (e.g., Granvik et al. 
2017). To link this to large impactors, they had to use the crater counts on older terrains, because craters 
produced by large projectiles are rare in the recent Solar System history. To calculate the recent impact 
flux of large asteroids from this exercise, the knowledge of the time dependence of the impactor flux is 
required. 

But here it is not expected that the time dependence of crater production rate over the past $\sim$4 Gyr was 
the same for small and large impactors. This is because the dynamical delivery of impactors during the 
early epochs was controlled by {\it size-independent} dynamical processes (e.g., Nesvorn\'y et al. 2017).
At the present time, instead, the impactor delivery from the main belt is regulated by the {\it 
size-dependent} Yarkovsky effect (see, e.g., Strom et al. 2005).
Most small NEA impactors, for example, evolve from the main belt by drifting into the $\nu_6$ resonance 
(Bottke et al. 2002, Granvik et al. 2017). Consequently, they have higher impact probabilities (normalized 
to one object) on the terrestrial worlds than large asteroids, which reach the NEA orbits via different escape 
routes (see below).  

The time dependence (chronology) of the impactor flux in Neukum et al. (2001) was constructed from small 
craters/impactors and therefore has the Yarkovsky effect in it. It is not obvious how to extrapolate the 
current impact flux it predicts to large impactors, which are affected by the Yarkovsky effect to a lesser degree. 
It would also be strictly incorrect to use the counts of large craters on ancient lunar terrains and extrapolate
them to recent times with the time dependence of the production function extracted from small crater counts. 
In fact, the impactor flux of large asteroids on the terrestrial worlds must have dropped more steeply from 
$\sim 4$ Gyr ago to present than the Neukum production function would suggest. This is likely the source of 
the order-of-magnitude discrepancy discussed above. 
\section{Methods}
We used the Wide-field Infrared Survey Explorer (WISE) catalog (Mainzer et al. 2011) to select all known $D>10$ 
km main-belt asteroids, 8193 in total, with semimajor axes $2<a<3.5$ au. The known population with WISE-diameter 
measurement is essentially complete for $D>10$ km.
We only consider large asteroids, because: (1) the motivating problem of this work concerns the impact flux of 
large asteroids, (2) the known population of small main belt asteroids is incomplete and models based on it 
would suffer from various uncertainties, and (3) the dynamics of small asteroids is affected by the 
radiation torque known as the YORP effect (Vokrouhlick\'y et al. 2015), whose strength depends on small scale 
surface features (e.g., Statler 2009) and is difficult to model with confidence (e.g., Bottke et al. 2015). 

Three clones were considered for each selected asteroid. The first clone was given the maximal 
negative Yarkovsky drift rate, the second one was given the maximal positive Yarkovsky drift rate, 
and the third one was given no drift. The maximal/minimal Yarkovsky drift rates were assigned 
to each individual body depending on its size and semimajor axis (see, e.g., Bottke et al. 2006). The 
drift rates were kept fixed during each orbital integration. This is an adequate approximation because 
the strength of the YORP effect scales as $1/D^2$ and quickly drops for large $D$. The obliquity 
changes due to the (neglected) YORP cycles should therefore be modest. We considered clones with 
the maximal and minimal drift rates to test how the impact flux depends on the full range of possibilities. 
Bodies with intermediate drift rates are expected to lead to intermediate results. With three clones 
for each $D>10$ km asteroid with $2<a<3.5$ au, we have nearly 25,000 bodies in total. 

Our numerical integrations included planets, which were treated as massive bodies that 
gravitationally interact among themselves and affect the orbits of all other bodies, and asteroid clones, 
which were massless (i.e., they did not affect each other and the planets). The integrations were
performed with the {\it Swift} $N$-body program (Levison \& Duncan 1994), which is an efficient 
implementation of the symplectic Wisdom-Holman map (Wisdom \& Holman 1991). Specifically, we used the 
code known as {\tt swift\_rmvs} that we adapted for the problem in hand. First, the code was modified
such that it can be efficiently parallelized on a large number of CPUs. Second, we modified the treatment
of close encounters between planets and asteroid clones such that the evolution of planetary orbits 
on each CPU is strictly the same (this code is known as {\tt swift\_rmvs4}). Third, we included the 
Yarkovsky force in the kick part of the integrator.

The collisional evolution was ignored, because $D>10$-km asteroids in the main belt have very long 
collisional lifetimes ($>$1 Gyr, Bottke et al. 2005; the collisional lifetime is asteroid's mean 
lifetime against catastrophic disruption). We did not include the YORP effect in the integrations,
because the main effect of YORP should be to produce extreme obliquity values and maximize the Yarkovsky
drift, which is accounted for by considering clones with maximal and minimal drift rates (but see
Vokrouhlick\'y et al. 2003). The dependence of the results on the YORP cycles could be tested with the 
stochastic YORP code that has been calibrated on various datasets (Bottke et al. 2015), but this is left 
for future work. We do not expect any major differences, because the YORP cycles of $D>10$-km main-belt 
asteroids are very long.

The integrations were performed on NASA's Pleiades Supercomputer. The integration with 25,000 asteroid clones 
was split over 2,500 Pleiades cores with each core dealing with 10 clones. All planets except Mercury were 
included. Leaving out Mercury allowed us to perform the simulations with a reasonably low CPU cost. The 
gravitational effects of Mercury were found insignificant in previous works (e.g., Granvik et al. 2016). 
The integration required 120 hours on 500 Pleiades cores. It covered 1 Gyr allowing us to monitor the asteroid 
impacts on the terrestrial planets during this time. It was run forward from the current epoch such that 
the results obtained from it should be strictly applicable to the impact flux during the next 1 Gyr. 
Still, with some uncertainty, the impact flux obtained from our integration can be thought as being 
representative of that at the present epoch.

The orbital elements of asteroids escaping into NEA space were saved at fixed time intervals and used as 
an input for an \"Opik-style collisional code (e.g., Greenberg 1982, Bottke \& Greenberg 1993, Vokrouhlick\'y 
et al. 2012). The code was used to compute the impact flux on different terrestrial worlds (Venus, Earth, 
Mars and the Moon). The effects of gravitational focusing were included. The {\it Swift} integrator recorded 
all impacts of asteroid clones on the terrestrial planets that occurred during the integration. This offers 
an opportunity to compare the number of recorded impacts with the impact profiles obtained from the \"Opik code. 
The results based on the {\it Swift}-recorded impacts should be more reliable than the results obtained 
from the \"Opik code, but the statistics from the recorded impact is small (see below). We use the strength 
of each method to derive the most reliable estimates. 

We also performed several additional integrations: (1) with a 1-day step over 100 Myr, and (2) using 15 asteroid 
clones distributed between the maximum negative and maximum positive values of the Yarkovsky drift. The former 
simulation was used to validate the nominal simulation with a longer timestep (3 days). We have not detected any 
statistically significant differences in the impact rate, number of NEAs or any other parameter. Simulation (2) 
was used to confirm that using clones with intermediate drift rates does not change the results. The increased 
statistics with 15 clones also helps to produce a larger number of recorded impacts in {\it Swift} and guarantee 
that the statistics derived from the {\it Swift}-recorded is reliable.    
\section{Results}
\subsection{Escape rate of $D>10$ km asteroids}
\begin{figure}
\epsscale{0.463}
\plotone{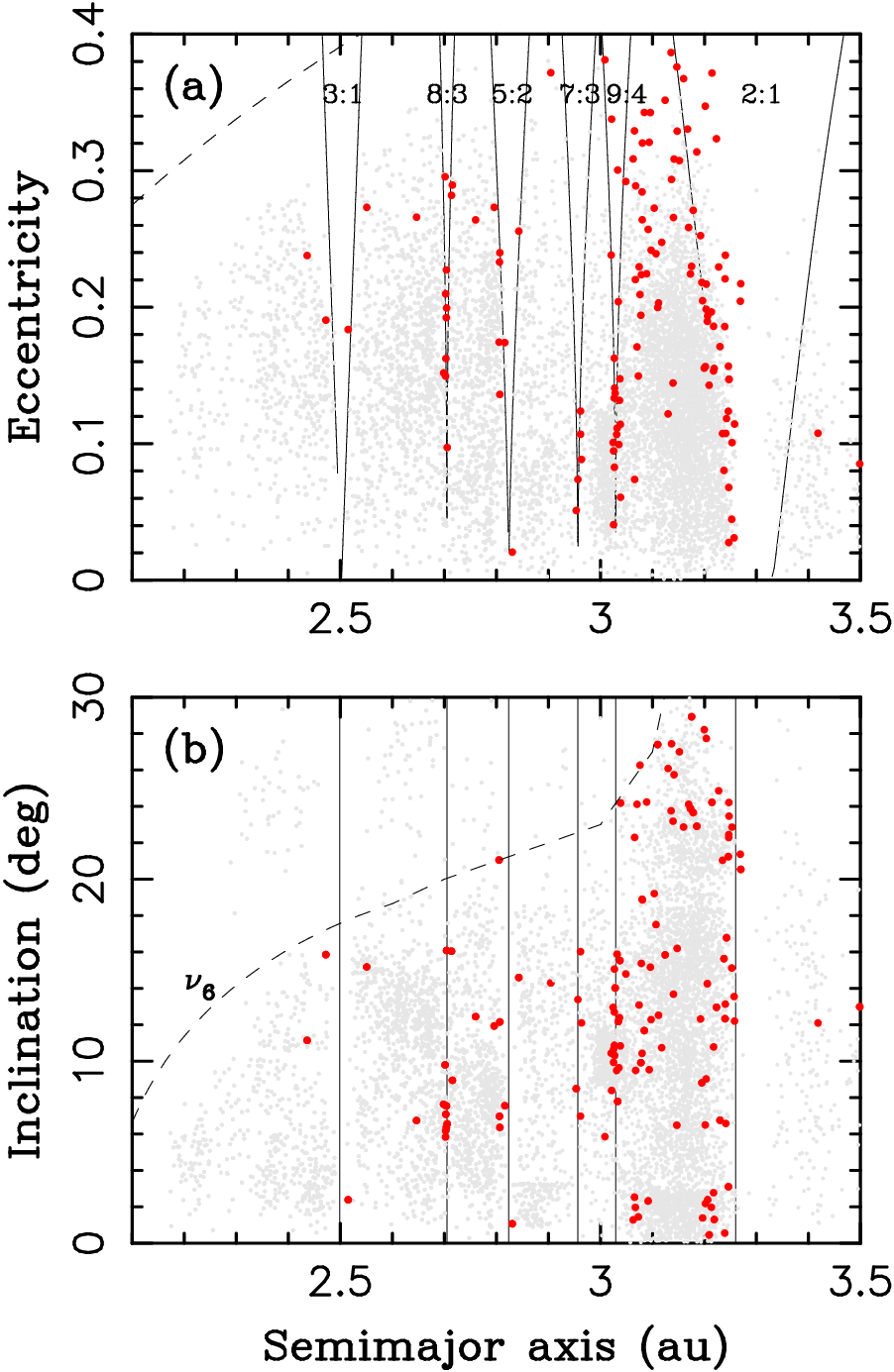}
\epsscale{0.45}
\plotone{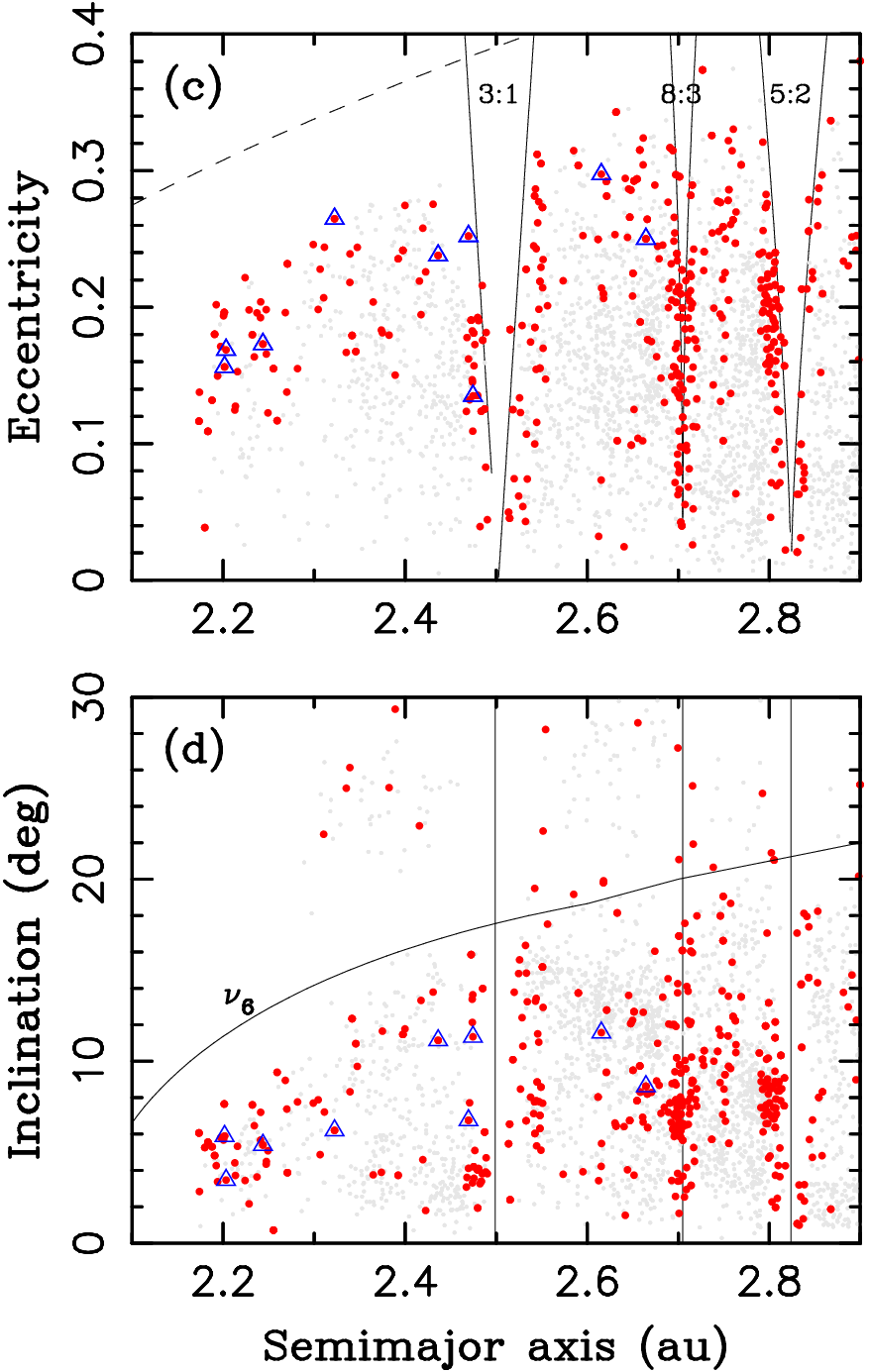}
\caption{Panels (a) and (b): The initial orbits of $D>10$ km main belt asteroids that remained in the main 
belt at 100 Myr after the start of our integration (gray dots) and those that escaped from the belt in 100 Myr 
(red dots). Only the clones with zero Yarkovsky drift were considered in the plot on the left. 
Panels (c) and (d): A zoom in on the inner part of the belt. Here we used the full length of the 
integration (1 Gyr) and all 15 clones. The triangles denote asteroids that ended up impacting on
the terrestrial planets. The thin solid lines denote the location of orbital resonances. The thin dashed 
line in (a) and (c) is the Mars crossing limit.}
\label{escape}
\end{figure}

Figure \ref{escape}a,b highlights the initial orbits of asteroids that escaped from the main belt 
in the course of our integration. This figure illustrates a couple of things. First, in what concerns the 
$D>10$~km bodies, the outer belt represents a much larger population than the inner belt. For example, 
when we compare the number of $D>10$ km asteroids with $a>2.9$~au and $a<2.9$ au, we find that the outer 
part of the belt contains more than twice as many asteroids than the inner part (about 
5,700 for $a>2.9$ au vs. about 2,500 for $a<2.9$~au). Second, many more large asteroids escaped from the outer
main belt than from the inner main belt: 114 bodies escaped from $a>2.9$ au in the first 100
Myr of the integration, while only 26 bodies escaped from $a<2.9$ au. Here we only counted clones with
zero Yarkovsky drift. The results for the maximum Yarkovsky rates were similar (e.g., 21 and 27 bodies escaped
from $a<2.9$ au with maximum negative and maximum positive drifts). Therefore, $\simeq5$ times
more large asteroids leak into NEA space from $a>2.9$~au then from $a<2.9$ au. The escape rates represent 2\% 
and 1\% of $a>2.9$ au and $a<2.9$ au populations per 100~Myr, respectively. 

Above we discussed the results from the first 100 Myr of our simulation, because the escape rates remain 
roughly constant during this time interval and should most closely represent the present situation. For $t>100$ 
Myr, the escape rate shows a slow decline such that the average escape rate over 1 Gyr is some $\simeq40$\%
lower than the one expected from the first 100 Myr. Perhaps the escape rate is really declining because we live 
in a special era when the escape rate is enhanced by some past event (e.g., Culler et al. 2000, Mazrouei et al. 
2017). The decline of the escape rate in our simulation may also be a consequence of some physical effect 
that we did not take into account (e.g., related to the rate and orientation of the Yarkovsky drift rates, 
collisional evolution). 

In total, only 1.7\% of bodies were eliminated from the belt in 100 Myr, indicating that the main belt is 
not losing much of its population at the present epoch. With 8192 $D>10$~km asteroids initially, the escaped population 
represents 139 $D>10$ km bodies in total. This means that one $D>10$ km asteroid leaves the main belt on average 
every 0.72~Myr. Since the escape rate determined from our simulation over 1 Gyr is some 40\% lower, if real, 
this would mean that the long term average is one $D>10$ km asteroid escaping from the main belt every 
1.2 Myr. 

Figure \ref{escape}c,d shows in more detail the inner part of the belt ($a<2.9$ au), which is 
important as for the impacts flux on the terrestrial planets, because the impact probabilities of
bodies evolving from this region are much higher that the ones evolving from $a>2.9$ au (e.g., Gladman 
et al. 1997). As for the orbital distribution of asteroids that were eliminated from $a<2.9$ au,
Figure \ref{escape}c,d indicates that most eliminated orbits started close to principal mean motion resonances 
with Jupiter, such as 3:1, 8:3 and 5:2. There is a prominent concentration of eliminated 
orbits on the left size of the 5:2 resonance with $e\sim0.2$ and $i\sim8^\circ$ corresponding to the 
Dora family (Family Identification Number or FIN 512 in Nesvorn\'y et al. 2015). 

As for the distribution of orbits eliminated from the innermost part of the belt ($a<2.5$~au), which is
the main source of impactors on the terrestrial worlds, Figure \ref{escape}c,d does not show any obvious 
concentration toward strong resonances ($\nu_6$ or 3:1). Instead, bodies are found to escape from the inner main via 
weak mean motion resonances with Mars and Jupiter (Migliorini et al. 1998, Morbidelli \& Nesvorn\'y 1999, 
Farinella \& Vokrouhlick\'y 1999). A small fraction of the escaping bodies started in the Flora family at 
the inner edge of the asteroid belt (FIN 402) and the Nysa-Polana complex on the left side 
of the 3:1 resonance (FIN 405). Overall, the contribution of families is small ($<$10\% of bodies 
reaching NEA space).

As we mentioned above, the escape rates of bodies with positive and negative Yarkovsky drift rates differ 
slightly. Considering the simulation with increased statistics (15 Yarkovsky clones and $a<2.9$ au), 
we find that the Yarkovsky effect acts to increase the escape rate of $D>10$ km from the main belt by $<$50\%, 
where the maximum difference corresponds to the maximum negative and maximum positive Yarkovsky drifts, 
relative to the case with no Yarkovsky drift. Therefore, if the spin vectors of $D>10$ km asteroids were 
randomly oriented in space, their escape rate from the belt would be only slightly higher than the average 
escape rate obtained with no Yarkovsky drift. The reality is more complicated, because the spin vectors of 
real asteroids show preferred orientations (Hanu\v{s} et al. 2011), which may correlate with the 
proximity to resonances.   

\subsection{The population of large NEAs}

Many of the escaping asteroids evolved onto the NEA orbits with perihelion distance $q<1.3$ au. We monitored 
the escaping bodies in a selected time interval $\Delta T$ and recorded the total time spent by asteroids 
on orbits with $q<1.3$ au, $T_{q<1.3}$, in $\Delta T$. To estimate the number of NEAs in a steady 
state, $N_{\rm NEA}$, we simply computed $N_{\rm NEA}=T_{q<1.3}/\Delta T$. Also, we extracted from the simulation 
the number individual bodies, $N_{q<1.3}$, that reached $q<1.3$~au during $\Delta T$. To estimate the mean 
dynamical lifetime on a NEA orbit, we computed $T_{\rm life}=T_{q<1.3}/N_{q<1.3}$. We then investigated how 
$N_{\rm NEA}$ and $T_{\rm life}$ depend on $\Delta T$, the source location of bodies (e.g., $a<2.9$ au or $a>2.9$ 
au), and the Yarkovsky drift rate assigned to them. 

We found that the inner ($a<2.9$ au) and outer ($a>2.9$ au) parts of the belt contribute to large NEAs in
a similar proportion. For example, with $\Delta T=100$ Myr, which more closely corresponds to the present epoch,
the expected steady state number of $D>10$ km NEAs produced from $a<2.9$ au is $N_{\rm NEA}=0.42$, while
$N_{\rm NEA}=0.57$ from $a>2.9$ au. As we pointed out above, $\sim5$ times more
asteroids escape from the outer part than from the inner part, but this factor is compensated in terms
of the contribution to the NEA population,
because the mean lifetime in NEA space of asteroids escaping from the outer belt is much shorter, $T_{\rm life}=0.52$ 
Myr, than that of the fugitives from $a<2.9$ au ($T_{\rm life}=1.6$ Myr). Therefore, each of many bodies escaping
from $a>2.9$ au typically lives shorter in NEA space, and therefore contributes less to the NEA population 
as a whole.

Interestingly, the results obtained with $\Delta T=1$ Gyr show slightly higher values of $N_{\rm NEA}$ than 
those obtained with $\Delta T=100$ Myr. For example, for the set of asteroids starting with $a<2.9$ au,
$N_{\rm NEA}$ obtained with $\Delta T=1$ Gyr is some 20\% higher than $N_{\rm NEA}$ obtained with 
$\Delta T=100$ Myr. This is a trend opposite to that seen in the escape rate, where the average over
1 Gyr was lower that the one over 100 Myr. We carefully analyzed the results to explain the source
of these differences and found that the asteroids that escape in the first 100~Myr are more concentrated 
toward outer mean motion resonances (5:2 and 7:3) and show shorter $T_{\rm life}$ in the NEA space after 
reaching NEA orbits. In fact, with $\Delta T=1$ Gyr and $a<2.9$ au, we obtain $T_{\rm life}=3.0$ Myr, 
an almost 2 times higher value than for $\Delta T=100$ Myr, because many bodies escaping late start 
with $a<2.5$ au and have longer $T_{\rm life}$. More work will be needed to establish with confidence whether 
these trends are real and, if so, what is their meaning. 

There is also some difference in $N_{\rm NEA}$ for different Yarkovsky clones. For example, for $\Delta T=1$ Gyr 
and $a<2.9$ au, we get $N_{\rm NEA}=0.41$ to 0.73 when different Yarkovsky clone sets are considered,
with the maximum values occurring for maximum negative and maximum positive drift rates. This roughly 
represents a 30\% uncertainty. The Yarkovsky effect contribution to the dynamical origin of $D>10$ km
NEAs in therefore not large.

We used the methods described above to obtain the total number of $D>10$ km NEAs expected in a steady state.
We found that the long term average ($\Delta T=1$ Gyr) is $1.2\pm0.4$, where the uncertainty is mainly related to
unknown Yarkovsky drift rates of individual bodies. Abstracting from details, we thus conclude that our 
best estimate with a generous error is $1.1\pm0.5$ NEA with $D>10$ km in a steady state. 

This estimate can be compared with the number of real $D>10$ km NEAs. The largest known NEAs are 
(1036) Ganymed ($D\simeq37.7$ km from WISE) and (433) Eros (mean $D\simeq16.8$~km from the NEAR 
imaging).  In addition, (3552) Don Quixote has $D\simeq26.7$~km (from WISE) and $a=4.23$ au and is 
very likely cometary in nature. The NEAs (1627) Ivar and (4954) Eric have $D\simeq8.4$ km and $D\simeq8.3$ km,
and are therefore smaller than our diameter cutoff. In summary, there are only 2 NEAs with $D>10$ 
km, while our model suggests $1.1\pm0.5$. The current population of $D>10$ km NEAs is thus 
$\sim1.3$-3.3 times larger than expected from our model. 

\begin{figure}
\epsscale{0.6}
\plotone{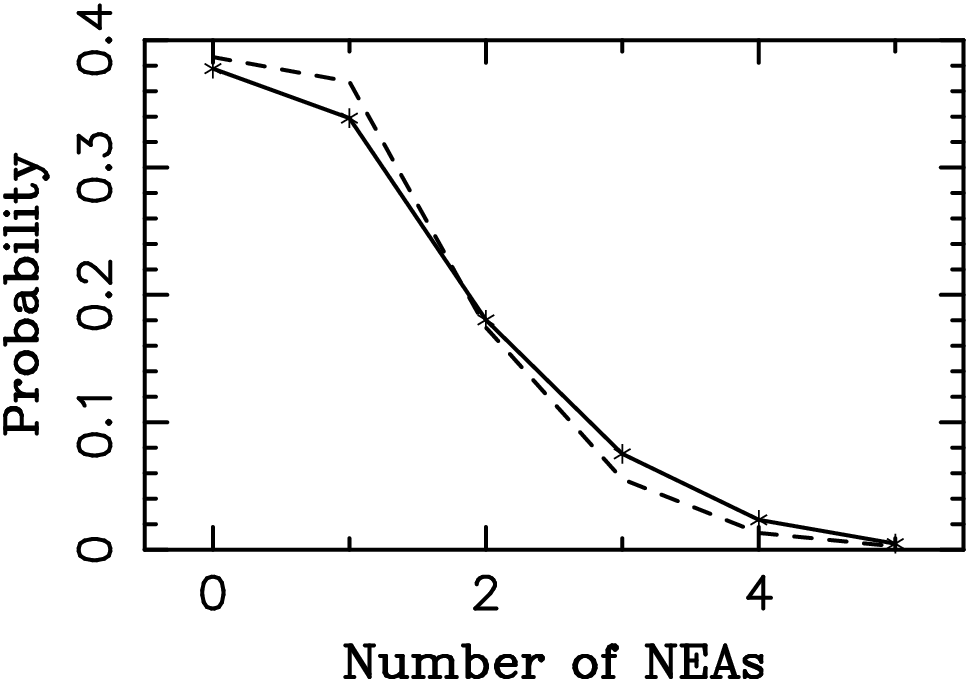}
\caption{The model probability of having a specific number of $D>10$ km asteroids in NEA space at the present
epoch (asterisks and solid line). The dashed line shows the best fit Poisson distribution with the 0.95 occurence 
rate.}
\label{prob}
\end{figure}

This appears to be related to statistical fluctuations in the number of large NEAs with the present population 
being slightly larger than the average. To test this, we used our simulation to compute the number 
of NEAs at different epochs (Figure \ref{prob}). We found that having 0, 1, 2 or 3 $D>10$ km NEAs occurs with 
probabilities 0.38, 0.34, 0.18 and 0.08, respectively (the remaining 0.02 corresponds to having more than 3). 
It is thus not unusual to have two $D>10$ km NEAs at the present time. The probability distribution 
can be fit by the Poisson distribution $P(n)=e^{-\lambda} \lambda^n/n!$, where $n \geq 0$ and the occurence 
rate $\lambda\simeq0.95$ (dashed line in Figure \ref{prob}).

\begin{figure}
\epsscale{0.7}
\plotone{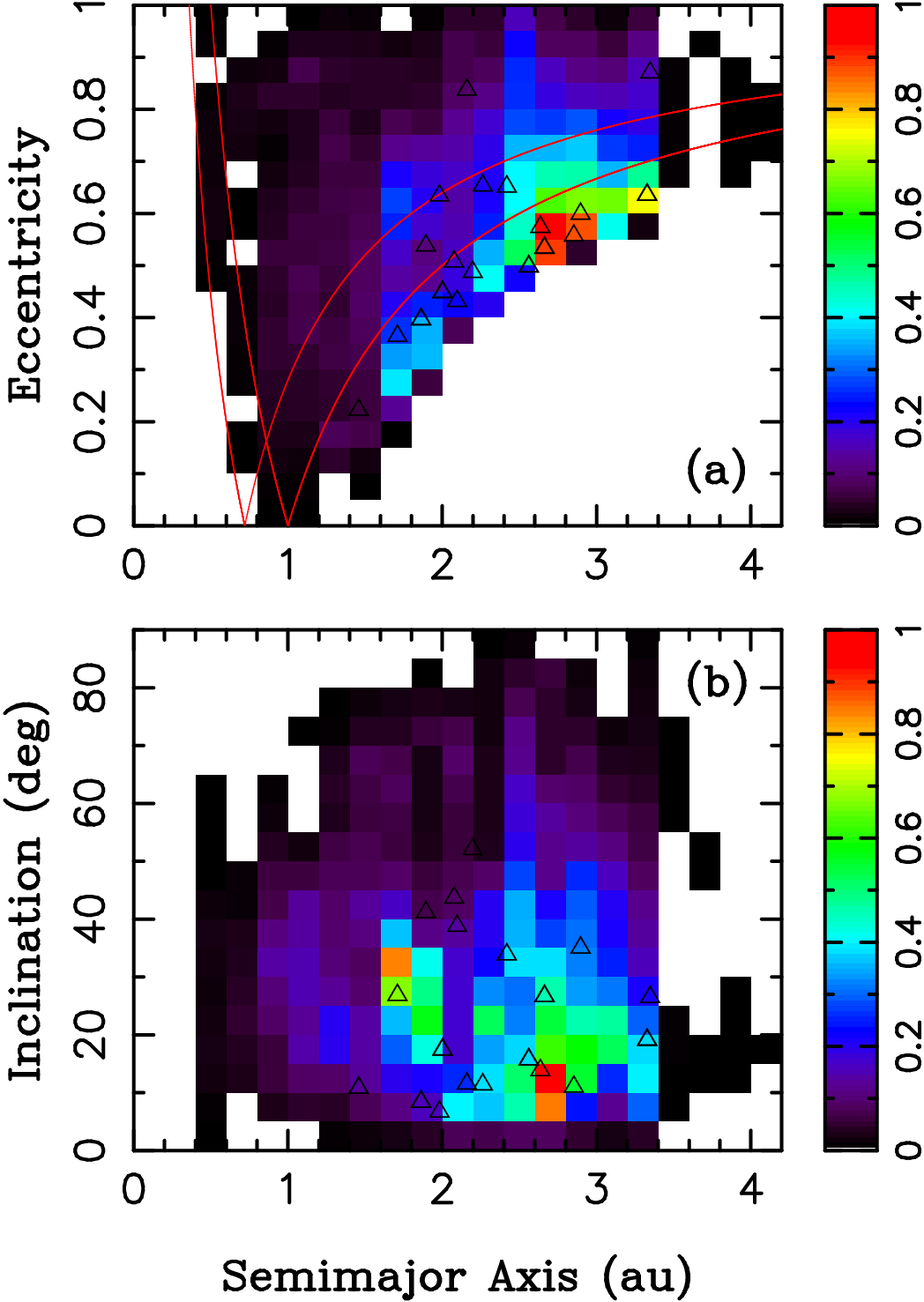}
\caption{The orbital distribution of $D>10$ km NEAs obtained in our model. The color scale, which expresses
the binned probability, appears on the right. 
The triangles show 22 known NEAs with $D>5$ km (data from WISE). We plot $D>5$ km NEAs here, 
because there are not enough 
$D>10$ km NEAs to make any sensible comparison between the model and observations. Also, our additional 
integrations with $D>5$ km asteroids show that they have similar dynamics to $D>10$ km asteroids and 
thus represent a useful reference.}
\label{resid}
\end{figure}

We used our model results to determine the steady-state orbital distribution of large asteroids 
in NEA space. Figure \ref{resid} shows the distribution of $D>10$ km bodies in a plot where we computed the 
residence time of model NEAs on different orbits and binned them in $(a,e)$ and $(a,i)$ projections. Compared 
to Bottke et al. (2002), who presented similar plots applicable to smaller NEAs (mainly $D\sim1$ km),
our orbital distribution seems to be more concentrated at $2.5<a<3.0$ au with fewer NEA orbits below
2.5 au. This may correspond to the tendency of large NEAs to sample the inner and outer parts of the asteroid
belt more equally than small NEAs (which predominantly start in the inner belt, roughly below~2.5 au).

The orbital distribution shown in Figure \ref{resid} does not have a very good statistics
and more work will be needed to test things. It is also not clear whether the orbital distribution of large
NEAs is consistent with the model expectation. This is because there are only two $D>10$ km NEAs to which our model
strictly applies. As a proxy, we plotted in Figure \ref{resid} all known 
NEAs with $D>5$ km. Their orbital distribution appears to be similar to the model-derived distribution 
for $D>10$ km. Using $a=2.5$ au to split the NEA population into two parts, for example, we find that there should be 
roughly the same number of large NEAs with $a<2.5$ au and $a>2.5$ au. For comparison, there are 12
$D>5$~km NEAs with $a<2.5$ au and 10 with $a>2.5$ au. 

\subsection{Impact flux of $D>10$ km NEAs}

The results of our numerical integrations were analyzed to resolve the motivating problem 
discussed in Section 2. From the number of impacts recorded by the {\it Swift} integrator and from 
the \"Opik code we determined that $0.8\pm0.3$ impacts of $D>10$ km asteroids are expected on the Earth 
in 1 Gyr. The estimated impact flux is thus well below the NEA-derived estimate of $\sim$7 $D>10$-km impacts 
per Gyr (Johnson et al. (2016), see also Chapman \& Morrison (1994), Stuart \& Binzel (2004), Le Feuvre 
\& Wieczorek (2011) and Harris \& D'Abramov (2015)). This confirms our expectations from Minton \& Malhotra (2010) 
and Nesvorn\'y et al. (2017). It shows that it is difficult to correctly estimate the steady-state impact 
flux of large asteroids from NEAs, because there are only a few large asteroids in NEA space at the 
present time, and because the collision probability of small NEAs cannot be extrapolated to large sizes. 

Specifically, Johnson et al. (2016) assumed that there are 5 NEAs with $D>10$ km, including 
(1627) Ivar and (4954) Eric, which in fact are just over 8 km according to WISE. (3552) Don Quixote is 
probably a nearly extinct comet on a wide orbit and has a low terrestrial impact probability. 
This brings the number of relevant NEAs down to two (Ganymed and Eros). 

In addition, if we interpret things correctly, the long term average of the number
of $D>10$ km NEAs is $1.1\pm0.5$ with the present population being a slight fluctuation above the average. 
Moreover, Johnson et al. (2016) applied the mean collisional probability with the Earth of $1.5\times10^{-3}$ 
Myr$^{-1}$ for each NEA (e.g., Stuart 2001). This is appropriate for small NEAs, from which this estimate was
derived, but not for large NEAs. We calculate that the mean collisional probability for each NEA
with $D>10$ km is only $\simeq 7\times10^{-4}$ Myr$^{-1}$, i.e. about two times lower. This happens 
because the outer part of the main belt is an important source of large NEAs (with about a half of their  
population deriving from $>2.9$ au). These outer belt asteroids have low terrestrial impact probability 
and weight down the mean.
 
In total, in the Gyr-long simulation with 15 clones, there were 38 impacts recorded on the terrestrial 
worlds (17 on Venus, 13 on Earth, and 8 on Mars; here we put together all our simulations for $D>10$ km
to improve statistics). About 70\% of the impactors started with $a<2.5$ au.
This shows that the inner belt is the dominant source of impactors. The remaining impactors started 
with $2.5<a<2.8$ au (i.e. between the 3:1 and 5:2 resonances). None of the impacting particles had 
$a>2.8$ au initially, which demonstrates that the contribution of the main belt with $a>2.8$ au to 
the terrestrial impacts is 
relatively small. With the \"Opik code we find that 70\% of all terrestrial impactors started with $a<2.5$ 
au and 30\% of impactors started with $2.5<a<2.8$ au. The Flora family is identified as the most 
important individual source of large impactors ($\simeq$10\% of the total recorded impacts).

As for impacts on different target worlds, we found from the \"Opik code that 0.84, 0.82, 0.42 and 0.044 
$D>10$ km impacts are expected over 1 Gyr on Venus, Earth, Mars and the Moon, respectively. The impact 
rates derived from recorded impacts is similar (1.1, 0.9 and 0.5 impacts on Venus, Earth, Mars).
Venus and Earth thus see comparable impact fluxes of large asteroids. Mars and the Moon receive roughly 
1/2 and 1/20 of the number of Earth impactors, which is consistent with the previous work (e.g. Nesvorn\'y 
et al. 2017). We found that the impact flux of $D>10$ km asteroids does not depend much on the assumed 
Yarkovsky drift; the population of clones with zero, negative and positive Yarkovsky drift rates all 
give the average number of 0.6-1 impacts on the Earth in 1 Gyr. 
\section{Conclusions}
Existing estimates of the current flux of planetary impactors in the inner Solar System vary by a factor 
of $\sim$10 in different publications. For example, it has been suggested from the analysis of NEAs that 
$\simeq$7 impacts of diameter $D>10$ km bodies should have occurred on the Earth in the last 1 Gyr 
(e.g., Chapman \& Morrison 1994, Stuart \& Binzel 2004, Le Feuvre \& Wieczorek 2011, Harris \& D'Abramov 
2015, Johnson et al. 2016). The impact rates inferred from models of NEA delivery from the main 
belt are at $\sim$7-14 times lower (e.g., Minton \& Malhotra 2010, Nesvorn\'y et al. 2017). Here we show 
that this problem is caused, in part, by the extrapolation of the impact flux from small ($D\sim1$ km) to
large sizes ($D>10$~km), where it is commonly assumed that the orbital distributions of small and large NEAs 
are the same. 

In reality, they are not the same because small and large NEAs reach their orbits by different 
dynamical pathways. To demonstrate this, we conducted dynamical simulations of $D>10$ km main-belt asteroids 
as they evolve onto NEA orbits by radiation forces and resonances. We found that the current impact flux of 
$D>10$ km asteroids on the Earth is $0.8\pm0.3$ Gyr$^{-1}$. The average impact probability of a $D>10$-km 
NEA is $\simeq$2 times lower than that of a $D\sim1$-km NEA. Additional differences arise because:
(1) most NEA-based studies used the absolute magnitudes to estimate that there are 5 $D>10$ km NEAs,
while they are only two (Ganymed and Eros; or 3 if Don Quixote is counted), and (2) because the current
number of two $D>10$ km NEAs may be a slight fluctuation above the long term average of $1.1\pm0.5$.  

Our work has important implications for our understanding of large impacts in the inner Solar System.
For example, the K/T extinction event at 65 Ma is thought to have been caused by an impact of $D\sim10$ km 
asteroid (Alvarez et al. 1980). Here we estimate that the average interval between terrestrial impacts 
of $D>10$ km asteroids is 1.2 Gyr. Having a $D\sim10$ km asteroid hitting the Earth just 65 Ma may thus 
be somewhat special. Alternatively, the impactor could have been smaller. We may have also missed in our 
modeling some important dynamical mechanism that facilitates escape of large asteroids from the main belt. 
More complete models will need to be developed to help to disperse these doubts.
  
\acknowledgments
This work was supported by NASA's SSERVI and Brazil's CNPq  ``Science without Frontiers'' programs. 
The simulations were performed on NASA's Pleiades Supercomputer.  We greatly appreciate the support of the 
NASA Advanced Supercomputing Division. We thank M. Delbo for kindly providing the catalog of sizes of the main 
belt asteroids. We thank W. F. Bottke and D. Vokrouhlick\'y for many helpful discussions. We also thank
David Minton for a helpful referee report.

\clearpage

\clearpage

\clearpage

\end{document}